\documentclass{emulateapj}

\usepackage{rotating}
\usepackage{natbib}
\usepackage{apjfonts}
\newcommand\tabspace{\noalign{\vspace*{0.7mm}}}
\def\errtwo#1#2#3{$#1^{+#2}_{-#3}$}

\newcommand\smc{SMC~X-1}

\newcommand\isis{{\tt ISIS}}

\newcommand\rxte{\textsl{RXTE}}

\newcommand\asm{\textsl{ASM}}

\newcommand\slang{{\tt S-lang}}
\newcommand\sitar{{\tt SITAR}}

\newcommand\aproxgt{\mathrel{%
      \rlap{\raise 0.511ex \hbox{$>$}}{\lower 0.511ex \hbox{$\sim$}}}}
\newcommand\aproxlt{\mathrel{%
      \rlap{\raise 0.511ex \hbox{$<$}}{\lower 0.511ex \hbox{$\sim$}}}}

\slugcomment{Submitted 2007, February 2.  Resubmitted 2007, June 18. Accepted 2007, July 19.}

\shorttitle{Tracking the Orbital and Super-orbital Periods of SMC~X-1}
\shortauthors{Trowbridge, Nowak, \& Wilms}

\begin{document}

\title{Tracking the Orbital and Super-orbital Periods of SMC~X-1}

\author{Sarah Trowbridge, Michael A. Nowak\altaffilmark{1}, J\"orn
Wilms\altaffilmark{2}} \altaffiltext{1}{Massachusetts Institute of
Technology, Kavli Institute for Astrophysics and Space Research,
Cambridge, MA 02139, USA; saraht@mit.edu, mnowak@space.mit.edu}
\altaffiltext{2}{Dr. Karl Remeis-Sternwarte, Astronomisches Institut der 
Universit\"at Erlangen-N\"urnberg, Sternwartstr. 7, 96049 Bamberg, Germany}

\begin{abstract}
The High Mass X-ray Binary (HMXB) SMC~X-1 demonstrates an orbital
variation of $\sim$3.89 days and a super-orbital variation with an
average length of $\sim$55 days. As we show here, however, the length of the
super-orbital cycle varies by almost a factor of two, even
across adjacent cycles. To study both the orbital and
super-orbital variation we utilize lightcurves from the \textsl{Rossi
X-ray Timing Explorer All Sky Monitor} (\rxte-\asm).  We employ the
orbital ephemeris from \citet{wojdowski:98a} to obtain the average
orbital profile, and we show that this profile exhibits complex
modulation during non-eclipse phases.  Additionally, a very
interesting ``bounceback'' in X-ray count rate is seen during
mid-orbital eclipse phases, with a softening of the emission during
these periods.  This bounceback has not been previously identified in
pointed observations.  We then define a super-orbital ephemeris (the phase
of the super-orbital cycle as a function of date) based
on the \asm\ lightcurve and analyze the trend and distribution of
super-orbital cycle lengths.  SMC~X-1 exhibits a bimodal distribution
of these lengths, similar to what has been observed in other systems
(e.g., Her~X-1), but with more dramatic changes in cycle length.
There is some hint, but not conclusive evidence, for a dependence of
the super-orbital cycle length upon the underlying orbital period, as
has been observed previously for Her~X-1 and Cyg~X-2.  Using our
super-orbital ephemeris we are also able to create an average
super-orbital profile over the 71 observed cycles, for which we
witness overall hardening of the spectrum during low count rate times.
We combine the orbital and super-orbital ephemerides to study the
correlation between the orbital and super-orbital variations in the
system, but find that the \asm\ lightcurve provides insufficient
statistics to draw any definitive conclusions on possible
correlations.
\end{abstract}


\section{Introduction}\label{sec:intro}

\setcounter{footnote}{0}

      SMC X-1, first discovered with \textsl{Uhuru} observations
\citep{leong:71a}, is a High Mass X-ray Binary (HMXB) consisting of a
neutron star \citep[an X-ray pulsar with 0.71\,s period;][]{lucke:76a}
and a young B0 supergiant companion \citep{webster:72b,liller:73a}.
An accretion disk is formed around the neutron star, likely partly via
wind-fed accretion where a strong stellar wind from the companion star
blows mass beyond its Roche Lobe radius and into the gravitational
influence of the neutron star.  Our view of this system is at high
inclination, as we witness X-ray eclipses of the neutron star and disk
by the companion star once every orbital period \citep{schreier:72a}.

These previous studies of the SMC~X-1 system measured the orbital
period to be approximately 3.892 days.  As has been observed in
similar HMXB systems, SMC X-1 also exhibits a long time scale
($\approx 60$\,days) super-orbital variation in its X-ray lightcurve
\citep{gruber:84a}.  Unlike the super-orbital variation in systems
such as Her X-1 \citep{tananbaum:72a}, which has a relatively
predictable 35 day period length \citep{staubert:83a}, the
super-orbital cycle length in SMC~X-1 is highly variable and follows
no obvious pattern \citep{gruber:84a,wojdowski:98a}.  As we elaborate
upon further below, lengths of the super-orbital cycles in SMC~X-1
can vary by up to a factor of two.

Accretion disk systems with super-orbital variations have been
explained with warped disks, seen close to edge-on such that the warp
partially obscures our view of the X-ray source \citep{katz:73a}.
Such warps are possibly due to an instability driven in the outer disk
by radiation from the central X-ray source
\citep{petterson:77a,pringle:96a,maloney:96a}, or a number of other
mechanisms \citep[see][for a review]{caproni:06a}.
\citet{wojdowski:98a} suggested that the super-orbital variation seen
in SMC~X-1 is indeed due to obscuration by such a warped disk, as
opposed to intrinsic flux variations, since the flux and spectrum
during orbital eclipse are fairly insensitive to whether the system is
in a `low' or `high' state of the super-orbital variation.

It also has been suggested that SMC~X-1 possibly has multiple warp
modes in its accretion disk in order to account for the wide variation
in super-orbital cycle length \citep{clarkson:03a}.  For radiatively
driven warps, theoretical studies show a number of possible
``branches'' of mode solutions that encompass both prograde and
retrograde, as well as stable and damped warps
\citep{wijers:99a,ogilvie:01a}. On the other hand, observational
evidence has been found in similar systems suggesting interaction
between the super-orbital variations and the orbital period or some
other underlying `fundamental clock' \citep{boyd:04a}.  Such systems
have been shown, in some cases, to have super-orbital cycle lengths
equal to integer or half-integer multiples of an underlying clock,
which in the case of Cyg~X-2 is the orbital period, while in the case
of LMC~X-3 and Cyg~X-3 is not simply related to any known dynamical
period in the system \citep{boyd:04a}.  (The super-orbital variations
in LMC~X-3, however, are likely due to flux variations rather than
obscuration; \citealt{wilms:01a}.)  In Her X-1 the super-orbital cycle
length mainly exhibits three values randomly, each differing only by
half the orbital period \citep{staubert:83a,still:04a,klochkov:06a}.

In this work we use data from the \textsl{All Sky Monitor} (\asm) on
board the \textsl{Rossi X-ray Timing Explorer} (\rxte) to make a
comprehensive study of the long term X-ray light curve of SMC~X-1.  We
study and characterize both the average orbital variation and
super-orbital variations in this system.  The outline of our paper is
as follows.  In \S\ref{sec:data} we discuss the extraction and
reduction of the \asm\ data.  We first use these data to characterize
the average X-ray properties of the orbital period
(\S\ref{sec:orb}). Next, in \S\ref{sec:sorb} we discuss how we define
the ephemeris for the super-orbital variations, and then in
\S\ref{sec:fold} we present their average X-ray properties. In
\S\ref{sec:combo} we consider jointly the average X-ray properties of
the orbital and super-orbital variations.  Finally, we summarize our
conclusions and make comparisons to other systems, such as Her~X-1, in
\S\ref{sec:sum}.

\section{ASM Data Reduction}\label{sec:data}

The \asm\ consists of three scanning shadow cameras, each equipped
with a position sensitive proportional counter that views the sky
through a set of slits to measure the relative intensities and
positions of X-ray sources in the sky \citep{levine:96a}.  The
detector operates by comparing observations of intensities in any
given area of the sky to a
catalog of known X-ray sources to obtain a lightcurve (on time scales
as short as 90\,s) for each target in the field of view.  The \asm\
records data in three different energy channels.  Channels one, two
and three are sensitive to X-rays of energy 1.5--3\,keV, 3--5\,keV and
5--12\,keV, respectively.

We obtained definitive lightcurves for SMC X-1 from the \asm\ source
catalog at the NASA \rxte\ website\footnote{\tt
http://heasarc.nasa.gov/docs/xte/ASM/sources.html}, and analyzed data
beginning MJD 50088.4 and ending 54083.8.  As mentioned previously,
the \asm\ data consists of a least-squares fit to the sky's modeled
X-ray spectrum for all the known sources within its field of view.
For this reason the count rate (cps) measurements from the satellite
will occasionally fall below zero when the target source counts are
lower than the difference between the expected background and the
actual background; we have not excluded such negative counts from our
extracted lightcurves.

Our analysis was performed using {\tt ISIS} version 1.4.2-5
\citep{houck:00a}, using custom scripts written in {\tt
S-lang}\footnote{{\tt www.s-lang.org}} (the scripting language
embedded within \isis), as well as routines publicly available from
the {\tt S-lang/ISIS Timing Analysis Routines}\footnote{\tt
http://space.mit.edu/CXC/analysis/SITAR} (\sitar).  Using these
routines to read the \asm\ data, we only retained data points where
the \asm\ solution had a maximum $\chi^2$-value of $<1.5$.  Before
beginning any analysis, we performed a barycenter correction on the
lightcurve to account for the movement of the satellite
within the Earth-Sun system.

\begin{figure}
\epsscale{1}
\plotone{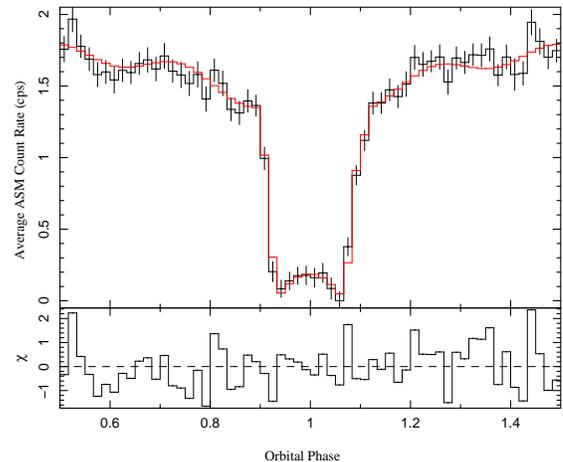}
\caption{The \asm\ lightcurve of SMC~X-1 folded on the orbital period,
fit with a 10 parameter empirical model.  Error bars are $1\,\sigma$, 
and are derived from the standard deviation of the mean rate in each bin. 
The bounceback feature is clearly visible between phases .875 and 
1.025\label{fig:orbital}}
\end{figure}

The \isis\ {\tt define\_counts} function allows arbitrary arrays of
lo/hi bin, value per bin, and error on that value to be registered as
a fittable data set, with presumed diagonal response matrix and unity
effective area.  We used this functionality to register the folded
\asm\ lightcurves of \S\ref{sec:orb} and \S\ref{sec:sorb} for fitting,
and used combinations\footnote{As opposed to {\tt IDL}, {\tt MATLAB},
or \isis's own array-based fitting -- where each unique model must be
a `separate' function -- \isis\ histogram fits use \slang\ to parse
model expressions such that almost any `mathematically sensible'
combination of individual model components results in a properly
defined total model. (This functionality will soon be extended to the
\isis\ array-based fitting as well; J. Houck, priv. comm.)} of custom
defined \slang\ functions to fit these data.
When it was necessary to fit a function to $(x,y)$ pairs rather than
histogram data sets (see \S\ref{sec:sorb}), we used the \isis\
function {\tt fit\_array}, which takes as input arrays of $x$- and
$y$-values, the relative weights of the data points, parameter initial
values and limits, and a reference to a single user-defined \slang\
fit function.

\section{Folding the Orbital Period}\label{sec:orb}

Our first step in characterizing the behavior of SMC X-1 was to search
the \asm\ lightcurve for evidence of the aforementioned periodicities.
Using a Lomb-Scargle periodogram \citep{lomb:76a,scargle:82a} we were
able to detect both a 55 day period (the average length of the
super-orbital cycles; see \S\ref{sec:sorb}) as well as a period of
approximately 3.89 days (i.e., the orbital period).  These findings
correspond to previous observations of the source
\citep{wojdowski:98a,wen:06a}.  We obtained an orbital ephemeris for
the source from \citet{wojdowski:98a}, which was defined in terms of
the initial observation time, the period and the period first
derivative.  Using this ephemeris with the \sitar\ routine {\tt
sitar\_pfold\_rate}, we folded the \asm\ lightcurve into 60 phase
bins, as shown in Fig.~\ref{fig:orbital}.

\begin{figure*}
\epsscale{1.2}
\plotone{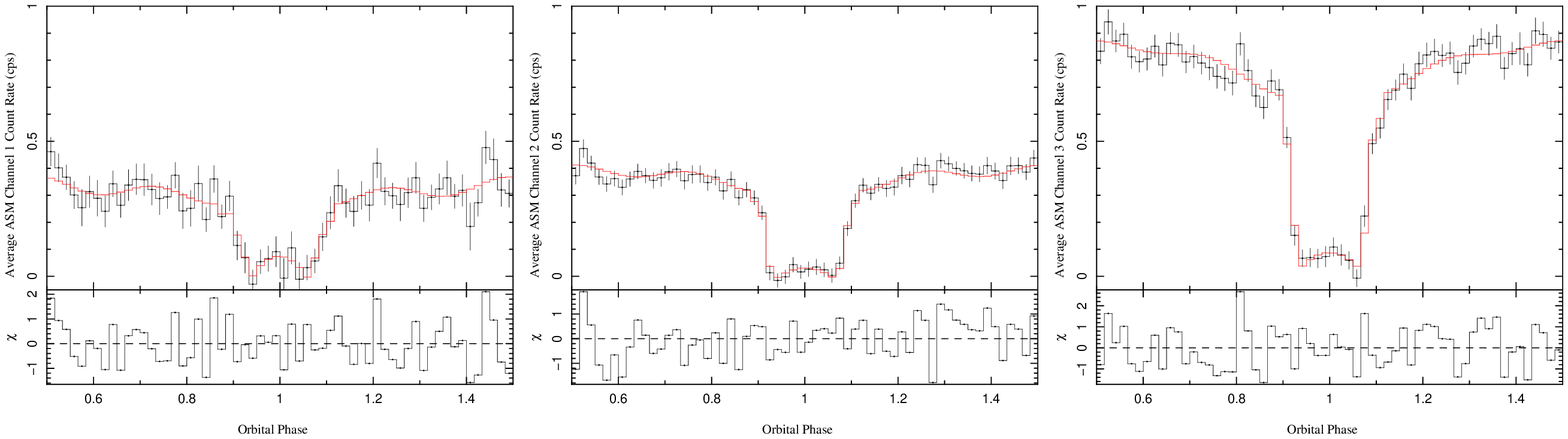}
\caption{Folded orbital profiles of the three \asm\ energy channels,
from left to right 1.5--3, 3--5 and 5--12\,keV.  Both the bounceback and
complex orbital modulation are more prominent in the lower energy
channels.
\label{fig:orbital_channels}}
\end{figure*} 

We had originally thought to search for improvements on the Wojdowski
et al.  ephemeris by detecting a drift in the zero point of orbital
phase (even relative to the known orbital period derivative from that
work) over the 11 years of \asm\ data we analyzed, but found that any
possible drift in the ephemeris was well within the observational
errors of the \asm. The \citet{wojdowski:98a} ephemeris remains accurate.

      Using the folded lightcurve, we were able to observe several
features of the orbital profile.  The most interesting of these
features is a ``bounceback'' phenomenon occurring during the eclipse.
Because we are observing the system at high inclination, we expect to
have few to no counts during the eclipse phases, when the companion
star obscures the line of sight between the satellite and the neutron
star-disk system.  Surprisingly, we observe the low counts expected on
either side of the eclipse, but witness a small rise and fall of
counts during the mid-eclipse phases.  We will henceforth refer to
this region of recovering count rate as the ``bounceback.''  Another
interesting feature is a complex modulation, modeled below as
sinusoidal, that occurs throughout the signal profile, and that is
very visible during the non-eclipse phases.

         We were able to obtain a fit to describe these features of
the orbital profile using a ten-parameter empirical function.  Our
final model was obtained as follows.  We modeled the bounceback
feature in the center of the eclipse with a parabola.  Allowing a free
parameter for the magnitude of the curvature, which could range to any
real value (positive or negative), we noted that the curvature was fit
to positive values in all channels.  Eclipse ingress and egress were
modeled by linear functions, dropping from/rising to a constant value
outside of eclipse.  We multiplied the entire function by a sine wave
plus a constant to account for the modulation observed in the profile,
and subtracted a Gaussian from the center to account for a broader
component of the eclipse, possibly due to partial obscuration of
potentially extended X-ray emission by the companion star and/or
accretion stream.  The center of the Gaussian and parabola were fixed
at 1, while the sinewave frequency was fixed at an integer or
half-integer multiple of the inverse orbital period.  We attempted the
fit with several different integer and half-integer values and used
the frequency with the lowest $\chi^2$ in our final fit function.  It
is interesting to note that the best fit was obtained modeling the
modulation with a sine wave of frequency 4 (in units of the inverse
orbital period), implying a complex disk warp structure.

Aside from the amplitude of the constant, amplitude and phase of the
sine wave, and area and standard deviation of the Gaussian, free
parameters also included widths for each of the regions: the eclipse,
the ingress, and the egress.  The folded data, fit and residuals are
plotted in Fig.~\ref{fig:orbital}, and the parameters are given in
Table~1.  This model, as well as other non-standard models used for
fitting in this paper, were written by the authors and can be made
available upon request.

  Once a successful model was obtained we were able to utilize it to
fit orbital profiles to each of the individual energy channels of the
\asm\ data (1.5--3, 3--5, and 5--12\,keV) and compare the
normalizations of features of the system in different energy ranges.
Fig.~\ref{fig:orbital_channels} shows profiles and fits in each of
these three energy channels, and the fit parameters are also presented
in Table~1.  In general we found that the lightcurve count rate was
dominated by the high energy portion of the spectrum.  As expected we
also found more modulation in the low energy channels than in the
high, likely due to preferential absorption of low energy X-rays, and
also possibly due to the relative sizes of the sources of high and low
energy X-rays.

Due to count rate uncertainties, it is not initially clear whether or
not the bounceback and sinusoidal variation are significant, real
features of the orbital profile.  In order to further explore the
significance of the bounceback, we performed fits of the orbital
profile without the parabolic term, resulting in a $\Delta\chi^2$ from
the fits to the full model described above of 14.1 in the summed
channel fit and 1.0, 2.0 and 10.0 in fits of the first, second and
third channels respectively.  We also performed the orbital profile
fits without the sinusoidal modulation, which resulted in a
$\Delta\chi^2$ of 13.2 in the summed channel fit and 5.4, 7.9 and 2.3
in the fits of the first second and third channels respectively from
the $\chi^2$-values obtained using the full model.  The statistics are
not good enough to allow study of the bounceback or the sinusoidal
modulation in individual periods.  The bounceback is not clearly
exhibited in averaging of most short portions of the lightcurve (i.e.,
only a few orbital periods) but can only be found with the statistics
provided by long averaging times.  A more thorough understanding of
the bounceback would require better statistics than presently
available from the \asm\ data.

\begin{deluxetable*}{cccccccccccc}
\tablefontsize{\scriptsize}  
\tablewidth{0pt}
\tablecaption{Orbital Profile: Constant*((1.0+Sinewave)*Orbit+Gauss)}  
\tablehead{ \multicolumn{1}{c}{ASM}   
            & \multicolumn{1}{c}{$\chi^2_{red}$}
            & \multicolumn{1}{c}{Constant}
            & \multicolumn{2}{c}{Sinewave} & \multicolumn{5}{c}{Orbit}
            & \multicolumn{2}{c}{Gauss}\\
\tabspace
            Channel  &   (50 DoF)
            & 
            & \colhead{Norm}
            & \colhead{Phase} & \colhead{Lwidth} 
            & \colhead{Rwidth}
            & \colhead{Cwidth} & \colhead{Height} 
            & \colhead{PNorm}
            & \colhead{Area} 
            & \colhead{Sigma} \\ }

\startdata
%
\tabspace
1+2+3  & 1.00 & \errtwo{1.74}{0.02}{0.02}
               & \errtwo{0.036}{0.016}{0.014} & \errtwo{0.72}{0.07}{0.06}  
               & \errtwo{0.100}{0.001}{0.005} & \errtwo{0.103}{0.005}{0.003}
               & \errtwo{0.134}{0.001}{0.001}
               & \errtwo{0.77}{0.02}{0.01} & \errtwo{0.10}{0.03}{0.03} 
               & \errtwo{-6.4}{0.4}{0.4} & \errtwo{0.18}{0.01}{0.02} 
               \\
\tabspace
1      & 0.95 & \errtwo{0.40}{0.05}{0.07} 
               & \errtwo{0.063}{0.038}{0.047} & \errtwo{0.68}{0.07}{0.06} 
               & \errtwo{0.117}{0.017}{0.015} & \errtwo{0.112}{0.013}{0.019} 
               & \errtwo{0.122}{0.878}{0.025}
               & \errtwo{0.73}{0.21}{0.33} & \errtwo{0.17}{0.17}{0.47}
               & \errtwo{-17.5}{7.5}{2.5} & \errtwo{0.41}{0.25}{0.21}
               \\ 
\tabspace
2      & 0.78 & \errtwo{0.39}{.04}{.01}
               & \errtwo{0.048}{0.026}{0.027} & \errtwo{0.78}{0.09}{0.09}
               & \errtwo{0.100}{0.006}{0.015} & \errtwo{0.106}{0.007}{0.018}
               & \errtwo{0.140}{0.860}{0.009}
               & \errtwo{0.87}{0.11}{0.18} & \errtwo{0.12}{0.14}{0.09}
               & \errtwo{-4.9}{1.8}{2.3} & \errtwo{0.17}{0.17}{0.06}
               \\
\tabspace
3      & 1.14 & \errtwo{0.86}{0.03}{0.01}
               & \errtwo{0.018}{0.018}{0.018} & \errtwo{0.74}{0.26}{0.74}
               & \errtwo{0.095}{0.002}{0.004} & \errtwo{0.100}{0.006}{0.001}
               & \errtwo{0.133}{0.007}{0.005}
               & \errtwo{0.74}{0.06}{0.08} & \errtwo{0.09}{0.07}{0.07}
               & \errtwo{-6.3}{1.2}{2.1} & \errtwo{0.17}{0.06}{0.04}
               \\
\enddata

\tablecomments{{\tt ORBIT} function parameters: Lwidth=width of the
eclipse ingress, Cwidth=width of the eclipse from low point on the
left of the bounceback to low point on the right of the bounceback,
Rwidth=width of eclipse egress.  Height=Distance from lowest point in
the eclipse to the constant in non-eclipse phases.  PNorm=Height of
the parabola fit in the bounceback (center of the eclipse) region.  In
Lwidth and Rwidth regions the profile was modeled with a linear
function and in the Cwidth region the profile was modelled as a
parabola.  Center of Orbit model and Gauss were frozen at phase=1 and
the sinewave frequency was frozen at f=4~(Orbital
Period)$^{-1}$. Error bars are 90\% confidence levels.}

\end{deluxetable*}

\begin{figure}
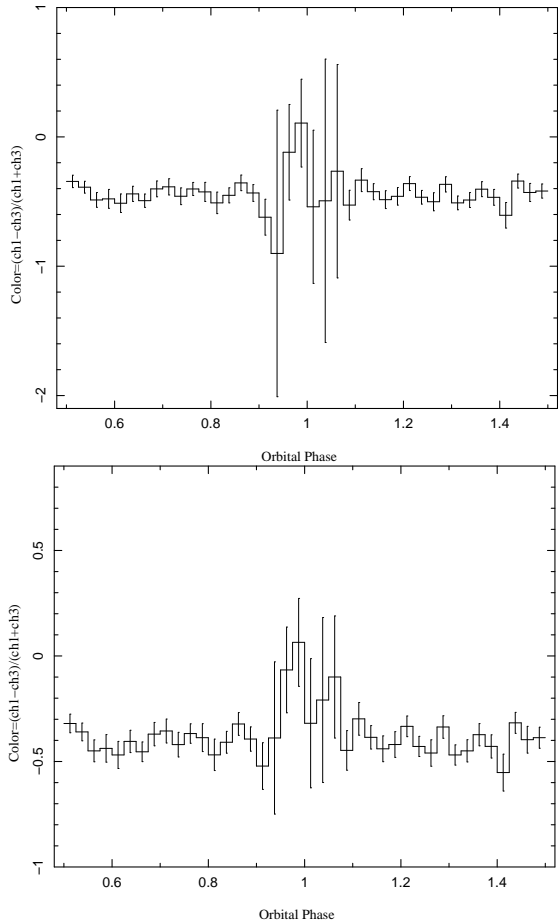

\epsscale{1}
\plotone{final_orb_color.ps}
\plotone{final_orb_color_shift.ps}
\caption{Top: The orbital-folded color lightcurve, exhibiting overall
hard X-ray dominated emission with softer X-ray emission in the
bounceback region. Error bars are are $1\,\sigma$ error bars. Bottom:
The same folded lightcurve as above; however, each energy channel
lightcurve was systematically shifted upward by 0.05\,cps.
\label{fig:color_orbital}}
\end{figure} 

In order to get a more quantitative idea of the energy distribution of
the emission, we created a folded color lightcurve, where we defined the
color, $C$, as
\begin{equation}
C=\frac{{\rm ch}1-{\rm ch}3}{{\rm ch}1+{\rm ch}3} 
\end{equation}\label{eq:color}
Here ${\rm ch}1$ is the count rate measured in \asm\ channel one, and ${\rm ch}3$
is the count rate measured in channel three.  Thus if the emission at
a certain phase is dominated by hard X-rays, then the value of the
color will be negative, and if the emission is dominated by soft
X-rays the color will be positive.  The orbital phase-dependent color
lightcurve (with 40 phase bins) is plotted in
Fig.~\ref{fig:color_orbital}.

Overall the color lightcurve demonstrates that the source is
relatively monochromatic and dominated, as observed in the channel
fits, by hard X-ray emission.  We did find that at the edges of the
eclipse, where the source was being partially obscured by the
companion star, that the X-ray emission possibly becomes much harder.
This might be attributable to soft X-ray absorption from the outer
atmosphere of the companion star; however, this region of the
lightcurve is also where the average ${\rm ch}1$ lightcurve dips below 0.
If we apply a systematic shift upward of 0.05\,cps to each of the
lightcurves, this hardening disappears, although the hardening remains
a possibility within the uncertainties.  On the other hand, the
bounceback emission, in contrast to the emission in the rest of the
orbital cycle, clearly becomes softer.

\section{Defining the Super-Orbital Ephemeris}\label{sec:sorb}  

     The inherent noise of the lightcurve and the highly variable
nature of the super-orbital cycle length provided somewhat of a
challenge as far as the definition of the super-orbital ephemeris (super-
orbital phase to date correlation) was
concerned. Here we wish to consider only variations on the longer
super-orbital time scale, and not those produced by the orbital
variations.  We therefore removed points that fell within the eclipse
of the neutron star by the companion from the lightcurve.  Using our orbital 
fold, discussed in the previous
section, we determined the orbital phases of the eclipse to lie
between 0.85 and 1.15 (Fig.~\ref{fig:orbital}), and used the
\citet{wojdowski:98a} ephemeris to calculate the time ranges during
which these phases would occur.  We then removed all \asm\ data points
that fell within those ranges.  

\begin{figure}
\epsscale{1}
\plotone{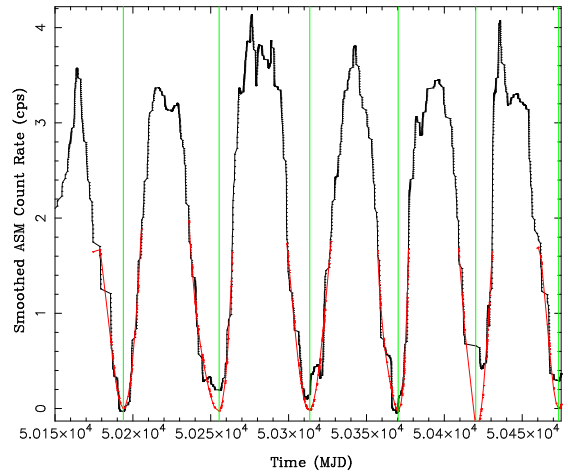}
\caption{SMCX-1 lightcurve between MJD 50150 and 50475, smoothed with
a Gaussian filter.  Fourth order polynomial fits in the minimum
regions are superimposed.  Minima of fits taken as the zero phase
points for the super orbital period are denoted by vertical lines.
\label{fig:so_fits}}
\end{figure}

Using the lightcurve with the eclipse phases removed, the \asm\
lightcurve remains intrinsically noisy.  For example, the peaks of
each super-orbital cycle have a number of prominent dips (see
Fig.~\ref{fig:so_fits}), which appear to be akin to the ``pre-eclipse
dips'' seen in the long term X-ray lightcurves of Her~X-1
\citep{shakura:98a,stelzer:99a,klochkov:06a}.  To further reduce the
inherent noise in the \asm\ lightcurves of \smc, we applied a simple
Gaussian Smoothing to the unbinned \asm\ data via an
FFT\footnote{Although only strictly valid for an evenly spaced
lightcurve, this procedure created a well-behaved lightcurve,
comparable to various schemes with wweighted averages, or binned
schemes with padding for data gaps, that we also tried.} with a
smoothing scale of 16 data bins ($\sim 2.5$\,days on average).  We
defined low count regions by selecting regions of the lightcurve that
were below a cutoff count value ($\le$1.675\,cps), and that were wide
enough ($\ge$ 7\,days) to assure that they were not residual noise
fluctuations from the lightcurve.  We then fit fourth order
polynomials to each of the low count regions using the \isis\ function
{\tt array\_fit}, choosing a uniform weighting of the data points (see
\S\ref{sec:data}).  We completed the definition of the super-orbital
ephemeris by defining the minima of the polynomials in the low count
regions as points of zero phase.  This procedure is illustrated in
Fig.~\ref{fig:so_fits}, which shows a portion of the smoothed
lightcurve with the polynomials and minima (zero-phase points)
superimposed.

\begin{deluxetable*}{ccccccccc}
\tablefontsize{\scriptsize} 
\tablewidth{0pt}
\tablecaption{Super-Orbital Zero Phase Times and Lengths}
\tablehead{\colhead{Cycle Number} & \colhead{Start Time} 
          &\colhead{Length}& \colhead{Cycle Number} 
          &\colhead{Start Time}& \colhead{Length}
          &\colhead{Cycle Number} & \colhead{Start Time}
          &\colhead{Length} \\ 
           \colhead{} & \colhead{(MJD)} & \colhead{(Days)} &
           \colhead{} & \colhead{(MJD)} & \colhead{(Days)} &
           \colhead{} & \colhead{(MJD)} & \colhead{(Days)} \\ }

\startdata

 1 & 50193.8 & 61.6 & 25 & 51420.4 & 59.6 & 49 & 52824.5 & 48.6 \\
 2 & 50255.4 & 58.2 & 26 & 51479.9 & 69.7 & 50 & 52873.2 & 48.6 \\
 3 & 50313.6 & 56.5 & 27 & 51549.6 & 61.0 & 51 & 52921.8 & 47.8 \\
 4 & 50370.1 & 50.0 & 28 & 51610.6 & 63.0 & 52 & 52969.6 & 61.4 \\
 5 & 50420.1 & 53.2 & 29 & 51673.6 & 70.1 & 53 & 53031.0 & 52.5 \\
 6 & 50473.4 & 46.8 & 30 & 51743.7 & 57.1 & 54 & 53083.5 & 52.7 \\
 7 & 50520.2 & 52.0 & 31 & 51800.8 & 56.8 & 55 & 53136.2 & 54.0 \\
 8 & 50572.2 & 45.2 & 32 & 51857.5 & 48.8 & 56 & 53190.1 & 62.3 \\
 9 & 50617.4 & 47.3 & 33 & 51906.3 & 58.6 & 57 & 53252.4 & 59.6 \\
 10 & 50664.7 & 40.3 & 34 & 51964.9 & 54.8 & 58 & 53312.0 & 61.2 \\
 11 & 50705.0 & 43.0 & 35 & 52019.7 & 66.7 & 59 & 53373.2 & 46.5 \\
 12 & 50747.9 & 49.9 & 36 & 52086.4 & 48.8 & 60 & 53419.7 & 66.6 \\
 13 & 50797.9 & 39.6 & 37 & 52135.2 & 59.7 & 61 & 53486.3 & 62.7 \\
 14 & 50837.5 & 50.7 & 38 & 52194.9 & 50.6 & 62 & 53548.9 & 59.6 \\
 15 & 50888.2 & 44.3 & 39 & 52245.4 & 64.8 & 63 & 53608.5 & 62.2 \\
 16 & 50932.4 & 49.4 & 40 & 52310.3 & 62.3 & 64 & 53670.7 & 54.8 \\
 17 & 50981.9 & 45.2 & 41 & 52372.5 & 46.8 & 65 & 53725.5 & 54.6 \\
 18 & 51027.0 & 65.7 & 42 & 52419.4 & 56.6 & 66 & 53780.1 & 54.4 \\
 19 & 51092.7 & 52.6 & 43 & 52476.0 & 72.1 & 67 & 53834.5 & 44.4 \\
 20 & 51145.3 & 57.9 & 44 & 52548.1 & 50.5 & 68 & 53878.9 & 47.1 \\
 21 & 51203.2 & 58.5 & 45 & 52598.5 & 49.1 & 69 & 53926.0 & 47.8 \\
 22 & 51261.8 & 52.3 & 46 & 52647.6 & 44.1 & 70 & 53973.8 & 42.2 \\
 23 & 51314.0 & 58.5 & 47 & 52691.7 & 71.9 & 71 & 54016.0 & 39.0 \\
 24 & 51372.5 & 47.8 & 48 & 52763.6 & 60.9 & 
\nodata & 
\nodata & 
\nodata \\

\enddata \tablecomments{We estimate our methods for determining zero
phase of the super-orbital cycles yields an uncertainty of about 2
days, which gives a corresponding uncertainty of $2\sqrt{2}$ days in
the determination of the super-orbital cycle lengths.}
\end{deluxetable*}

\begin{figure}
\epsscale{1.1}
\plotone{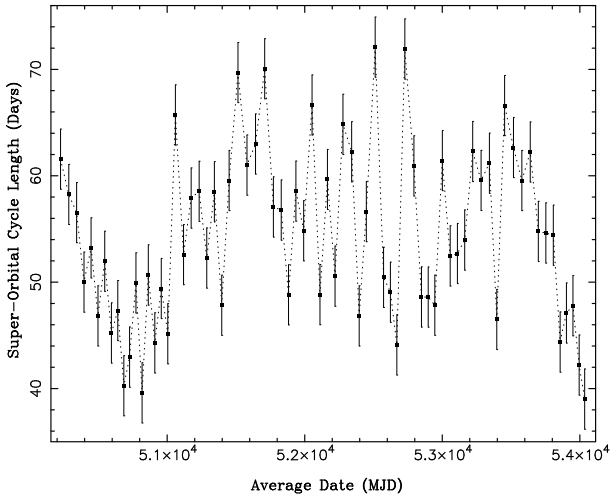}
\caption{The time evolution of super-orbital cycle length.  Error bars 
represent the uncertainty in cycle length due to the uncertainty in the
determination of super-orbital zero phase times.  The
oscillatory structure superimposed on the long-term trend has
previously been attributed to multiple warp modes in the disk
\protect{\citep{clarkson:03a}}. \label{fig:date_solength}}
\end{figure}

\begin{figure}
\epsscale{1}
\plotone{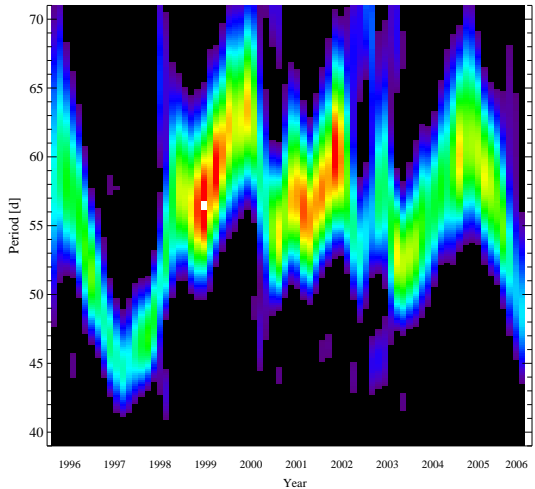}
\caption{Dynamical Lomb-Scargle periodogram for the entire \asm\
lightcurve of SMC~X-1.  The periodogram is calculated in 315 day
intervals, slid 50 days each bin.  (See \protect{\citealt{wilms:01a}}.)
\label{fig:lomb_smooth}}
\end{figure}

     With the ephemeris defined as in Fig.~\ref{fig:so_fits}, we were
able to extract the start and stop time of each super-orbital cycle.
We estimate, based upon multiple trials with variants of our
zero-phase search routines (weighted averages, binned routines, to
define the lightcurve, and several different functional forms to fit
the minima) that our cycle start times have an accuracy of $\pm2$
days.  To search for possible patterns, we plotted the length of each
super-orbital cycle versus the average date during that cycle.  This
plot is shown in Fig.~\ref{fig:date_solength}.  There appears to be a
somewhat oscillatory trend, superimposed on a much longer time scale
pattern.  The longer term trends, especially the short (40--50\,day)
super-orbital cycles near the beginning of the \asm\ lightcurve, are
likely what have been identified as multiple warp modes in the disk
\citep[i.e.,][]{clarkson:03a}.  Specifically, if we apply a sliding
Lomb-Scargle periodogram to the \asm\ data (exactly as we have
previously shown with \asm\ lightcurves of LMC~X-3;
\citealt{wilms:01a}), the rather ragged pattern of
Fig.~\ref{fig:date_solength} appears more as a smooth evolution of
super-orbital period, as shown in
Fig.~\ref{fig:lomb_smooth}. Comparing to Fig.~\ref{fig:date_solength},
however, we see that the smoothness of this trend is partly an
artifact of the averaging process, and that a sliding Lomb-Scargle
periodogram by itself is unrevealing for the fine detail of the
evolution of this source.

\begin{figure}
\epsscale{1}
\plotone{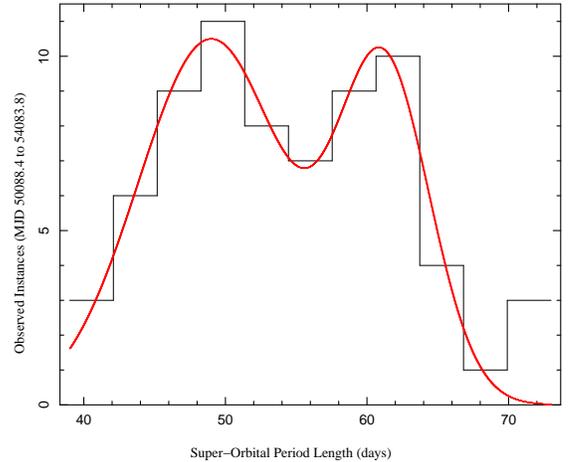}
\caption{Histogram distribution of super-orbital cycle lengths with
bimodal (sum of two Gaussians) fit superimposed.  The fit shown in
this figure was used in the simulations described in \S\ref{subsec:depend}.
\label{fig:so_dist}}
\end{figure}

We studied the distribution of these varying cycle lengths by
creating a histogram of their values.  The number of bins in the histogram 
was chosen because it produced the most easily seen pattern, but does not 
significantly change the results of the analysis.  For example, doubling the 
number of bins does not change the shape of the distribution, except to add 
more noise to the overall trend.  We found in this case that the
distribution has a distinct double-peaked pattern, similar to other
systems of this type.  Specifically, such systems as Cyg~X-2, LMC~X-3,
Cyg~X-3 and Her~X-1 show doubly peaked distributions for histograms of
wait times between successive minima \citep{boyd:04a}. It is possible
that the SMC~X-1 system is similarly oscillating (somewhat randomly)
between two extreme periods.  The profile of super-orbital cycle
lengths is shown in Fig.~\ref{fig:so_dist}.

     Once we had defined the ephemeris, we attempted to find some
correlation between the length of the super-orbital cycles and the
X-ray flux from the system.  Towards this end, for each super-orbital
cycle we binned the rates into a histogram and attempted fits with
several different functional forms; none of these were successful for
all of the super-orbital cycles.  This was due both to the large
variance in the shapes of each super-orbital variation (including the
aforementioned dips), as well as the lack of data for some portions of
the lightcurve.  Instead, we performed five point spline fits on each
of the super-orbital cycles. From the spline fits we extracted the
minimum, peak and mean flux during each super-orbital cycle
and plotted each of those values against both the super-orbital cycle
length, and the length minus its mean value of approximately 54.4 days.  
We were not able to find any correlations between super-orbital cycle length 
and flux with this method, nor by working on a coarser time scale, i.e., by
averaging over five consecutive super-orbital cycles and searching
for correlations between the same variables.

\section{Folding on the Super-Orbital Period}\label{sec:fold}

\begin{figure}
\epsscale{1}
\plotone{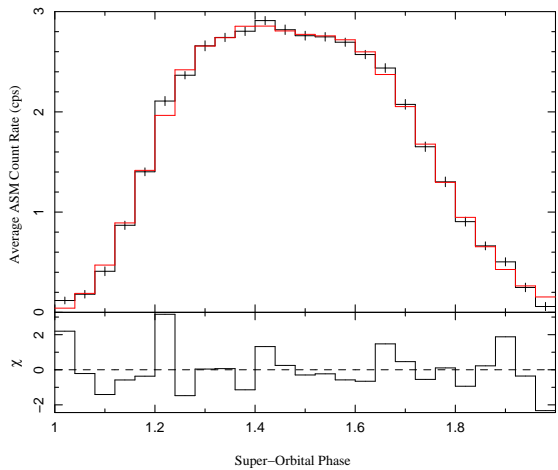}
\caption{Lightcurve Folded on super orbital period with fit of two 
superimposed Weibull functions.
\label{fig:super_orb}}
\end{figure}  

  With the super-orbital ephemeris defined, we were able to assign a
super-orbital phase to each point on the lightcurve and proceed to
fold the data over all super-orbital cycles.  We chose to average the
\asm\ lightcurve in 25 phase bins of the super-orbital cycles, and
show the results in Fig.~\ref{fig:super_orb}.  It is interesting to
note the asymmetry between the rise and fall of the curve, with the
leading edge (super-orbital eclipse egress) exhibiting a much sharper
rise than the falling edge's decline.  The best fit to this profile
was obtained with the sum of two scaled Weibull functions (i.e.,
$\propto x^\gamma \exp[-(x/\beta)^\alpha]$; see Appendix B of
\citealt{nowak:99b} for historical references and description of the
uses of this distribution function).  We elected to use Weibull
functions because their asymmetry can reasonably describe the
asymmetry of the folded lightcurve.  We used the sum of two curves
as this matched the broadly peaked nature of the folded profile
much better than did a single function.  This was also suggested by
the shape of the individual super-orbital cycle profiles, many of
which were dramatically double peaked.  Fit results are presented in Table 3.


\begin{deluxetable*}{cccccccccc}
\tablefontsize{\scriptsize} 
\tablewidth{0pt}
\tablecaption{Two Weibull Function Fits to Lightcurves Folded on Super-Orbital Cycle}
\tablehead{\colhead{ASM} & \colhead{$\chi^2_{red}$} 
          &\colhead{Norm$_1$} &\colhead{Peak$_1$} 
          &\colhead{$\alpha_1$}& \colhead{$\beta_1$} 
          &\colhead{Norm$_2$}& \colhead{Peak$_2$} 
          &\colhead{$\alpha_2$}& \colhead{$\beta_2$} \\
\tabspace
          \colhead{Channel} & (17 DoF) \\}
\startdata
1+2+3 & 1.95 & \errtwo{0.39}{0.08}{0.01} & \errtwo{1.28}{0.01}{0.00} 
             & \errtwo{2.34}{0.21}{0.02} & \errtwo{0.26}{0.02}{0.00}
             & \errtwo{1.30}{0.09}{0.10} & \errtwo{1.57}{0.01}{0.00}
             & \errtwo{4.76}{0.09}{0.55} & \errtwo{0.88}{0.09}{0.10}
\\
\tabspace
1     & 2.41 & \errtwo{0.04}{0.05}{0.02} & \errtwo{1.24}{0.03}{0.03} 
             & \errtwo{2.1}{3.3}{0.8} & \errtwo{0.16}{0.17}{0.07}
             & \errtwo{0.28}{0.03}{0.07} & \errtwo{1.52}{0.04}{0.03}
             & \errtwo{3.25}{1.81}{0.03} & \errtwo{0.63}{0.33}{0.23}
\\
\tabspace
2     & 2.07 & \errtwo{0.12}{0.13}{0.03} & \errtwo{1.30}{0.03}{0.01} 
             & \errtwo{2.5}{0.7}{0.4} & \errtwo{0.29}{0.06}{0.05}
             & \errtwo{0.27}{0.02}{0.03} & \errtwo{1.59}{0.01}{0.00}
             & \errtwo{6.1}{3.9}{1.9} & \errtwo{0.99}{0.31}{0.48}
\\
\tabspace
3     & 1.61 & \errtwo{0.15}{0.04}{0.01} & \errtwo{1.27}{0.01}{0.01} 
             & \errtwo{2.03}{0.15}{0.05} & \errtwo{0.20}{0.03}{0.01}
             & \errtwo{0.69}{0.00}{0.40} & \errtwo{1.54}{0.02}{0.01}
             & \errtwo{4.5}{0.9}{0.5} & \errtwo{0.86}{0.09}{0.09}
\\

\enddata \tablecomments{Weibull functions had the form Norm$_1 (x-x_0)
\exp[-((x-x_0)/\beta_1)^{\alpha_1}]$, etc., where $x_0$ was determined
by fixing the location of the function maximum to Peak$_1$.  Error
bars are 90\% confidence levels.}
\end{deluxetable*}

\begin{figure}
\epsscale{1}
\plotone{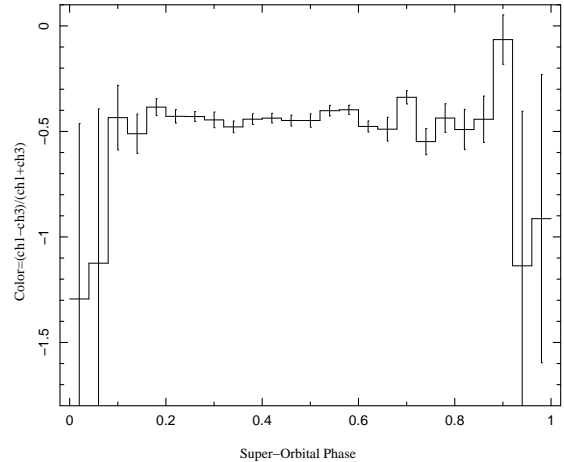}
\caption{Color diagram for super-orbital profile.  Emission is
relatively monochromatic and dominated by hard X-rays.  In lower
intensity regions the emission is harder than in high intensity
regions, which is in contrast to the eclipse emission in the average
orbital profile.
\label{fig:s_orbital_color}}
\end{figure}
  
     As in our analysis of the orbital profile, using the data from
the individual \asm\ energy channels we were able to create a
super-orbital profile for each of the three energy channels.  We fit
each of these profiles with the same functional form used to fit the total
profile.  Fit results for the individual channels are also presented in Table 3.
As expected from our analysis of the orbital profiles we once again
found that the profile was dominated by high energy emission.  Similar
to our analysis of the average energy profile of the orbital
variation, we created a color lightcurve for the super-orbital profile
by plotting the color versus super-orbital phase, again using the
definition of eq.~(1). We present these results in
Fig.~\ref{fig:s_orbital_color}.  Once again we can see that the
emission is fairly monochromatic, with two exceptions.  Near phase 0.9
of the super-orbital period, there is weak evidence for a softening.
More clear, however, is that near phase 0 of the super-orbital cycle
(i.e., the low count rate region in which the neutron star is likely
eclipsed by the disk) the emission appears to become harder, although
the color values in this region have relatively large uncertainty.
This is in marked contrast to the ``bounceback'' region of the orbital
period fold, where the emission became softer.  As in our color plot
for the orbital profile, we have examined the same plot with a
systematic shift upwards of 0.05 cps in all channels, however, in this
case it does not significantly change the result.

\section{Joint Orbital and Super-Orbital Profiles}\label{sec:combo}

\subsection{Two Dimensional Phase Folds}\label{subsec:twod}

\begin{figure}
\epsscale{1}
\plotone{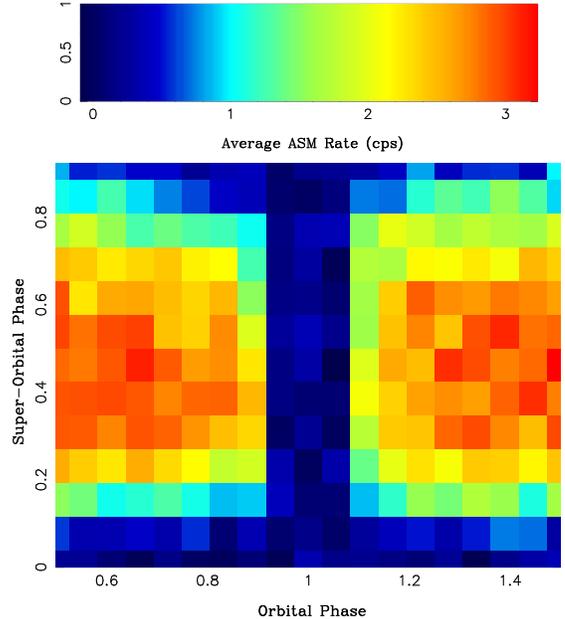}
\caption{Two dimensional folded lightcurve.  Variation in orbital
phase is plotted along the $x$-axis while super-orbital phase is
plotted along the $y$-axis.  The top panel shows the color scale.
\label{fig:td_fold}}
\end{figure}

   To better understand the link between the orbital and super-orbital
variations in SMC~X-1, we created a two dimensional fold of the
SMC~X-1 lightcurve.  Once we had defined an ephemeris for both the
orbital and the super-orbital variations (as discussed in
\S\ref{sec:orb} and \S\ref{sec:sorb}, respectively) we were able to
assign every point on the lightcurve both an orbital and a
super-orbital phase.  We then created a grid of histogram bins with
orbital phase on one axis and super-orbital phase on the other, and
sorted the data points into that grid, averaging the intensity
measurements in each bin.  The result was the two dimensional
histogram shown in Fig.~\ref{fig:td_fold}.  While we can observe the
asymmetry noted in the super-orbital profile in this histogram, it
does not contain sufficient statistics to give specific insight as to
the evolution of orbital profile features at different super-orbital
phases and vice versa.  The choice of binning is made for maximum clarity.  
Unfortunately, current statistics are not sufficient such that finer binning 
would provide more information.

\begin{figure}
\epsscale{1}
\plotone{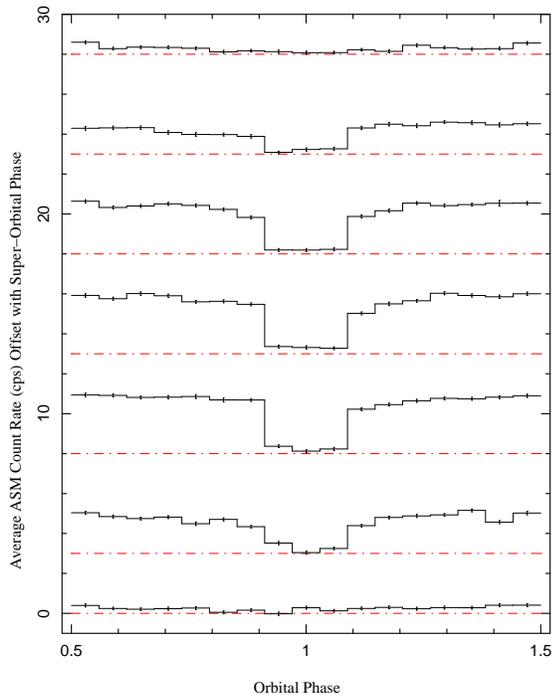}
\caption{Lightcurve folded on orbital period for specific ranges of super 
orbital phase, demonstrating the correlation between the two.  Statistics are 
not good enough to enable identification and analysis of the bounceback feature
 in these folds. \label{fig:smear_td_fold}}
\end{figure}

To gain another visualization of the form of the orbital profile at
different super-orbital phases, we also created several orbital
profiles, each containing only points of the lightcurve in a small
range of super-orbital phase, and plotted these profiles on a single
set of axes.  We present these orbital phase histograms in
Fig.~\ref{fig:smear_td_fold}.  Here again we find that we have
insufficient statistics to demonstrate any definitive trends in the
shape of the orbital profile dependent upon super-orbital phase.
Further study of this system with better statistics (e.g., with a
series of pointed X-ray observations) are likely necessary to witness any
trends.

\subsection{Searching for Dependencies upon Orbital Period}\label{subsec:depend}

     In accordance with patterns seen in other sources, e.g., Her~X-1
where the `turn on' of the super-orbital cycle is tied to a half
integer multiple of the orbital period
\citep{staubert:83a,klochkov:06a} or Cyg~X-2 where the time between
successive minima is an integer multiple of the orbital period
\citep{boyd:04a}, we searched for a correlation between the orbital
period length and the super-orbital cycle lengths.  We used a set of
$\sim$3500 values between one quarter of the orbital period and one
half of the longest observed super-orbital cycle to search for a
value that divided most evenly into all of the super-orbital cycle
lengths.  For any given member of the set, we tested its relation to
the super-orbital cycle lengths by dividing each of the 71 observed
cycle lengths by that value, and summing the
non-integer remainders of each of the 71 quotients.  A value that fit
relatively evenly into the set of super-orbital cycle lengths was one that
exhibited a minimum value of the sum of non-integer parts.  We found
16 values that exhibited minima in this formula, which we define as a
point for which the sum of non-integer parts was $\le$31.  (The
maximum value of the sum of the remainders was $\sim$43.6, while the mean of
the sum was $\sim$36.2.)  Of the 16 minima we found 7 that had some integer or
quarter-integer relation to the orbital period, i.e., approximately
0.25, 0.5, 1, 2, 3, 5 and 6 times the orbital period length.  The
other 9 values that exhibited minima had no obvious relation to the
orbital period length.

  To test the significance of these observed values, we created a
histogram of the sum of non-integer parts in order to study its
distribution.  The small value of this sum that we had obtained for
0.25, 0.5, 1, 2, 3, 5 and 6 times the orbital period length appeared
rare, but not necessarily unusually so; therefore, we performed a
Monte Carlo simulation to determine their significance.  We first fit
a bimodal (sum of two Gaussians) distribution to the super-orbital
cycle length data shown in Fig.~\ref{fig:so_dist}, from which we
could randomly choose super-orbital period lengths.  A single
simulation consisted of randomly choosing 71 super-orbital cycle lengths from
this distribution, and then performing the same analysis as done for the
actual data.  Specifically, the main goal was to create a histogram of
the sum of non-integer parts from the quotients of each of the 3500
trial values with the chosen set of super-orbital cycle lengths.  We
performed $10^6$ such simulations and averaged all of the resulting
histograms to obtain a theoretical distribution.

 From this simulated distribution we could estimate the likelihood of
obtaining the sums of non-integer remainders with values less than or
equal to those that had been found for the quarter-integer and integer
multiples of the orbital period.  We found these probabilities to be
$\approx$4.2\%, 0.1\%, 4.2\%, 3.6\%, 0.6\%, 4.2\% and 2.5\% for the
members of the set that were 0.25, 0.5, 1, 2, 3, 5 and 6 times the
orbital period length, respectively.  From these simulations we
concluded that while these values appear to have interesting relations
to the orbital period, and their corresponding sums of non-integer
remainders are somewhat rare, they are not rare enough to point
conclusively towards a relationship to the orbital period.  Further
study of this correlation -- either via \asm\ observations with a
longer baseline, or via pointed observations that can more accurately
determine the start and stop times of each super-orbital period -- may
prove useful in determining if there truly is a relationship between
the super-orbital cycle and orbital period lengths.

\section{Summary}\label{sec:sum}

In our analysis of the \asm\ lightcurve of SMC~X-1 we have observed
several interesting features of both the orbital and super-orbital
profiles.  By using the ephemeris provided in \citet{wojdowski:98a},
we were able to fold $\sim$11 years of data on orbital period to
obtain an average orbital profile. The main results of this analysis
are as follows.

\begin{itemize}

\item The lightcurve is characterized by complex modulation during the
persistent emission (here modeled as a sinusoid, but which could possibly fit 
another model within the uncertainty), with a broad, asymmetric eclipse.

\item We were able to witness a count rate bounceback phenomenon when averaging
 over large numbers of orbital cycles, which consists of a
small rise and fall in counts during orbital mid-eclipse phases.  

\item The emission becomes somewhat softer during the
eclipse/bounceback, while the persistent emission is fairly
monochromatic.

\end{itemize}

Despite several pointed observations during orbital eclipse, in both
low and high states of the super-orbital period, the bounceback has
not been previously seen \citep{woo:95a,wojdowski:98a}.  We did not
have sufficient statistics to observe the dependence of this
phenomenon on the super-orbital cycle, it is therefore possible that
the bounceback is only present in certain super-orbital phases, or
that its presence is only intermittent.  Given the relatively few
pointed observations of SMC~X-1 during eclipse, it is possible prior
pointed observations have simply not observed the state in which the
bounceback is most prominent.  

The softening during the eclipse might be akin to that seen in
so-called Accretion Disk Corona (ADC) sources \citep{white:82a}.  Such
sources show broad, partial eclipses, with deeper eclipses in the hard
X-ray \citep[for example, see the lightcurves of
X1822$-$371,][]{heinz:01a}.  One model of these sources has a very
spatially extended corona scattering X-ray from the central source
into our line of sight; however, X-rays traveling through the
atmosphere just above the midplane of the disk are preferentially
absorbed in the soft X-rays \citep{white:82a}.  Thus, the eclipse of
the base of this extended corona is preferentially blocking harder
X-rays, and leads to an overall softening of the spectrum.  However,
if such is occurring in SMC~X-1, this does not explain the
``bounceback'' of the lightcurve.  Perhaps additional scattering into
our line of sight at orbital phase 0, from the atmosphere of the
companion, rather than just an ADC, must also be invoked to explain
the bounceback.

The super-orbital variation in SMC~X-1 is not as easily dealt with as
the orbital variation because of the complicated nature of the
super-orbital ephemeris.  The super-orbital cycle length is highly
variable and Lomb-Scargle techniques alone are not sufficient to
define its structure \citep{wen:06a}.  In our analysis we were able to
obtain a super-orbital ephemeris from the \asm\ data by modeling the
locations of zero phase for each of 71 cycles.  From that ephemeris we
were able to extract an average super-orbital profile, as well as
information about trends in the super-orbital cycle length.  Our main
results are as follows:

\begin{itemize}

\item In general we observe an asymmetry in the average super-orbital
profile, most notable in a sharper rise in the low state egress than
fall in the low state ingress.

\item 
In contrast to the trend in the orbital profile, where we see softer
emission during low count states, we find that emission appears to harden 
during low states in the super-orbital profile.

\item We have found the lengths of the super-orbital cycles in
SMC~X-1 to be bimodally distributed, in agreement with
\citet{boyd:04a}, and have also found suggestive, although not
definitive, evidence that the variation in length is driven by some
internal clock, the fundamental period of which is related to the
orbital period as observed in Her~X-1 and Cyg~X-2. \citep{boyd:04a}.

\item In contrast to findings in analysis of Her~X-1
\citep{still:04a}, we have been unable to find any correlation between
the super-orbital cycle length, and X-ray flux.  

\item In contrast to \citet{clarkson:03a}, who report a smooth
evolution of the super-orbital cycle length, we observe sharp
variations, often with large changes in length between adjacent 
super-orbital cycles.  

\end{itemize}

This latter result is especially evident in observed super-orbital
cycles 41 through 44 (MJD 52372.5 to 52598.5), in which the cycle
duration rises from $\sim$47 days in cycle 41 to $\sim$72 days in
cycle 43, and falls back to $\sim$50 days in cycle 44.  Likewise
between super-orbital cycles 46 through 49 (MJD 52647.6 to 52873.2)
there is a rise from $\sim$44 days to $\sim$72 days in cycles 46 and
47, and a fall back to $\sim$49 days in cycle 49.
\citet{clarkson:03a} also theorizes that the evolution of
super-orbital cycle length is possibly sinusoidal.  The lack of a
regular, sinusoidal pattern in our data can easily be seen in
Figs.~\ref{fig:date_solength} and \ref{fig:lomb_smooth}.

In our spectral analysis of 71 super-orbital cycles we were able to
observe a clear super-orbital profile in the first channel
(1.2--3\,keV), of an amplitude similar to that of the modulation in
the second channel \citep[3--5\,keV; cf.][who report little or no
variation below 3\,keV]{clarkson:03a}.

The most likely reason for the discrepancies between our analysis and
that of \citet{clarkson:03a} is the extended length of the lightcurve
we were able to analyze (about twice the length of that used in
\citealt{clarkson:03a}) and our more precise definition of the edges
of each individual period.  The smoothing of super-orbital period
length variation they observed is most likely at least partially an
artifact of their use of a sliding Lomb-Scargle periodogram, and is
similar to that we observed in Fig. \ref{fig:lomb_smooth}.

The question then arises as to why \smc\ shows such extreme variations
in its super-oribtal cycle length, while sources such as Her~X-1 and
LMC~X-4 do not.  Several tentative suggestions have been made based
upon radiatively driven warp models. In such models, the driving force
of the warp is the flux from the central source, which scales as
$R^{-2}$, being absorbed and reradiated from the outer disk, whose
intercepting area scales as $R^2$, thereby yielding a torque that
increases linearly with distance.  Disks can therefore become unstable
to radiative driving of a warp if they are large enough \citep[and the
accretion efficiency is high enough;][]{pringle:96a}, which is a
function of the system's orbital period and secondary to primary mass
ratio.  Numerical simulations \citep{wijers:99a} and analytic
estimates \citep{ogilvie:01a} indicate that the Her X-1 system is just
large enough to have a stable warp.  The analytic models, however,
also indicate that the, rather complex in terms of their eigenvalue
behavior, underlying equations only admit stable warp solutions for a
finite range of disk sizes \citep{ogilvie:01a}.  Estimates are that
\smc\ may be near the upper limit of system sizes that are unstable to
warping, and therefore may be more likely to exhibit chaotic warp
modes.  Our results are certainly consistent with that expectation.

Within the framework of a radiatively driven warp, the existence and
stability of the warp modes are also dependent upon the radius at
which matter is injected into the disk \citep{wijers:99a,ogilvie:01a}.
Being more heavily wind-fed, this may be another way in which \smc\
differs from Her X-1.  Perhaps variations in the wind speed of the
\smc\ secondary, leading to variations of the disk circularization
radius (equated with the mass injection radius in the majority of
theoretical models) may account for the varying cycle lengths.  An
\asm\ study like the one presented here, coupled with spectroscopic
observations of the secondary's wind, may be an avenue for exploring
such a possibility.

The above speculations, however, do not account for the possibility of
a fundamental ``underlying clock'' for the \smc\ super-orbital cycles.
In their analysis of several X-ray binary systems \citet{boyd:04a}
find many systems for which the super-orbital excursion lengths are
random integer multiples of some fundamental period, (Cyg~X-2, LMC~X-3
and Cyg~X-3).  They also find, when plotting the lightcurves of the
systems in $X,\dot{X}$ phase space, that each of these sources
demonstrates circulation in the phase space about two rotation
centers, creating a bimodal distribution in the length of
super-orbital variations.  SMC~X-1 shows similar characteristics.

\citet{wijers:99a} speculated that for sufficiently large amplitude
warps, the effective mass injection radius (i.e., where the incoming
accretion flow interacts with and circularizes into the inclined disk)
could greatly fluctuate from one super-orbital cycle to the next,
causing fluctuations in cycle duration that become tied to the orbital
period.  An avenue for exploring this possibility may be to model
individual cycles of the \asm\ lightcurve, to see if one can discern a
changing warp amplitude with each cycle.  The fact, however, that we
did not find a dependence of cycle duration upon mean, minimum, or
peak \asm\ flux, may argue against this possibility.

To summarize, using the \asm\ lightcurve, we have defined a
super-orbital ephemeris for SMC~X-1 for 1995-2006.  It will now be
possible to put pointed observations during this period in the context
of both the orbital ephemeris from \citet{wojdowski:98a} and the
super-orbital ephemeris from this work.  We have exploited the
spectral information from the three channels of the \asm\ as much as
possible and extracted information from the spectrum in both the
orbital and super-orbital profiles, to guide both further
observations, as well as theories, e.g., of disk warping, which
currently do not describe behavior as complex as that seen here in
\smc.  It is also now necessary to perform multiple pointed
observations in order to observe a variety of super-orbital and
orbital phase combinations and gather better statistics than provided
by the data analyzed in this study.  It might then be possible to
analyze the super-orbital evolution of orbital profile features,
including the bounceback.  With pointed observations we might also be
able to better analyze the super-orbital ingress and egress and the
spectral variation in the emission, to obtain more insight into the
structure of the HMXB SMC~X-1 and other similar systems.

\acknowledgments This work was supported by NASA Grant SV3-73016.  The
authors gratefully acknowledge Al Levine for providing the code upon
which the \asm\ barycenter correction was based.  The authors would
also like to acknowledge help and advice from John Houck and Andy
Young. The authors thank the anonymous referee, whose comments helped
to improve the clarity of this paper.


\end{document}